\documentclass[prl,twocolumn,twoside,showpacs,superscriptaddress,floatfix]{revtex4}
\usepackage{graphicx,color,amssymb,amsmath,epsf}
\epsfclipon
\renewcommand\appendix{\parEq.
\setcounter{section}{0}
\setcounter{subsection}{0}
\setcounter{figure}{0}
\setcounter{table}{0}
\renewcommand\thesection{Appendix \Alph{section}}
\renewcommand\thefigure{\Alph{section}\arabic{figure}}
\renewcommand\thetable{\Alph{section}\arabic{table}}
}
\usepackage[colorinlistoftodos]{todonotes}

\usepackage{color} 

\begin{document}
\title{Analytic solutions for links and triangles distributions in finite Barab\'asi-Albert networks}
\author{Ricardo M. Ferreira}
\affiliation{Instituto de F\'\i sica\\ Universidade
Federal do Rio Grande do Sul\\ Av. Bento Gon\c{c}alves
9500, C.P. 15051 - 91501-970 Porto Alegre, RS, Brazil }
\author{Rita M. C. de Almeida}
\affiliation{Instituto de F\'\i sica\\ Universidade
Federal do Rio Grande do Sul\\ Av. Bento Gon\c{c}alves
9500, C.P. 15051 - 91501-970 Porto Alegre, RS, Brazil }
\affiliation{Instituto Nacional de Ci\^encia e Tecnologia
- Sistemas Complexos}
\author{Leonardo G. Brunnet}
\affiliation{Instituto de F\'\i sica\\ Universidade
Federal do Rio Grande do Sul\\ Av. Bento Gon\c{c}alves
9500, C.P. 15051 - 91501-970 Porto Alegre, RS, Brazil }
\date{\today}

\begin{abstract}
Barab\' asi-Albert model describes many different natural networks, often yielding sensible  explanations to  the subjacent dynamics. However, finite size effects may  prevent from  discerning  among different underlying physical mechanisms and from determining whether a particular finite system is driven by Barab\'asi-Albert dynamics. Here we propose master equations for the evolution of the degrees, links and triangles distributions, solve them  both analytically and by numerical iteration, and compare with numerical simulations. The analytic solutions for all these distributions predict the network evolution for systems as small as 100 nodes. The analytic method we developed is applicable for other classes of networks, representing a powerful tool to investigate the evolution of  natural networks.   
\end{abstract}


\maketitle

The pioneering works by Barab\'asi and Albert \cite{Barabasi1999a,
Barabasi1999b} propose a model that builds a network adding nodes and links with a preferential attachment mechanism, yielding a degree distribution following a power-law. Here degree is defined as the number of neighbors of a given node. Barab\'asi-Albert model was applied to a wide variety of networks such
as actor collaboration, WWW, power grid data and protein-protein
interaction (PPI) \cite{Albert2002,Barabasi2004}. It emerges as a possible representative of a wide class of natural systems. However, it is a common controversy whether a given system may or may not be reduced to a Barab\'asi-Albert network, with notorious examples given by metabolic pathways or PPI networks \cite{Przulj2004,Khanin2006,FoxKeller2005}. In fact, even networks simulated following Barab\'asi and Albert recipe may show deviations from the typical topological statistic measures due to finite size effects. In real data such deviations  may stem either from these finite size effects or from relevant differences in the underlying growth dynamics. In this context, analytic results are most welcome.

Our analytic characterization focuses on the dependence with time $t$ of
three distribution functions: the number of nodes of degree $k$, $N(k,t)$; the number of links between nodes with degree $k$ and $k'$, $L(k,k',t)$; and the number of triangles involving degrees $k$, $k'$ and $k''$, $\Delta(k,k',k'',t)$. The latter offers an insight of the neighborhood structure of nodes, whose common measure is the clustering coefficient defined as the number of 
triangles for which this node is a vertex, divided by the number of possible triangles formed by the node itself and two of its neighbors\cite{Watts1998}. Examples of clustering-dependent dynamics are cascading failures\cite{Ash2007} and epidemics \cite{Eguiluz2002,Newman2003}.

The only exact, analytic result for these three distributions when  $ t \rightarrow \infty$   is limited to  the power law $N(k,t) \propto k^{-3} $ for Barab\'asi-Albert and related models \cite{Barabasi1999a, Barabasi1999b, Dorogovtsev2000, Dorogovtsev2002, Krapivsky2000}. Further information on these distributions in the  $ t \rightarrow \infty$ limit is obtained only from simulations, as far as we know.

Nevertheless, network size is important. For example, infinite networks results may not discern between the effects due to the growth dynamics or the system finite size \cite{Albert2002,Dunne2002}. Additionally, epidemic outbreaks depend on network size \cite{Pastor-Satorras2002}, and  social\cite{Eguiluz2002} and protein association networks\cite{Fraser2005,Li2006,Ferreira2013} are typically smaller than 30000 nodes. For finite networks some approximated results are available in the literature. Fotouhi and Rabat \cite{Fotouhi2013} analytically obtained a solution for $N(k,t)$, valid for $k$ less than the initial number of nodes, $N_0$,  presenting a finite size correction and agreeing with previous results in the limit $ t \rightarrow \infty$. Also, the average clustering coefficient of a Barab\'asi-Albert network goes to zero for $t \rightarrow \infty$, but for finite time, the clustering coefficient for a given node was described by 
Fronczak \emph{et.al.} \cite{Fronczak2003}, using a mean-field approach, obtaining a dependence with $\ln(t)$ and time when the node was added. However, this result agrees with simulation data only for networks greater than $40000$ nodes \cite{Fronczak2003}. Finally, the total number of triangles in the networks goes as $\ln(t)$,  as obtained by Bianconi and Capocci \cite{Bianconi2003}, proving that the network clustering coefficient goes to zero for infinite networks.

In this work we analytically obtain the distributions functions $N(k,t)$, $L(k,k',t)$, and $\Delta(k,k',k'',t)$ as continuous time limit solutions of master equations and validate the results by numerically iterating these equations for discrete time and comparing to simulations of  $10000$ random networks generated following the usual Barab\'asi-Albert algorithm\cite{Barabasi1999a,
Barabasi1999b}. The analytic expressions are obtained  using a method that is valid for network sizes smaller than 100 nodes, enough time to fade out the initial condition transient. Furthermore, as we show below,  this method is straightforwardly extendable to  other algorithms provided that the corresponding master equations may be decoupled.

The master equations and solutions are obtained as follows.
We assume networks starting  with $N_0$ nodes, each having $k_0$ neighbors. 
For simplicity $k_0$ is assumed to be greater than $m$, the degree of
the new node added at each time step. Following Barab\'asi-Albert algorithm, the probability $\Pi(k,t)$ that a node with
degree $k$ receives a new connection is given by
\begin{equation}
\Pi(k,t)=\frac{m k}{\sum_{j=1}^{m_0+t-1} k_j}=\frac{m k}{2mt+N_0k_0} \; .
\end{equation}
The master equation for  the number of nodes with degree $k$, $N(k,t)$,
can be written as
\begin{eqnarray}
\nonumber N(k,t+1) &=& N(k,t) + \delta_{k,m} + \frac{m (k-1)}{(2mt +
N_0k_0)}N(k-1,t)\\
&-& \frac{m k}{(2mt + N_0k_0)}N(k,t) \; .
\label{master_N}
\end{eqnarray}
The second term on the right side of the equation represents the
creation of nodes with degree $m$, the third (fourth) term represents the
probability of a node with degree $k-1$ ($k$) to receive a link, increasing (decreasing) the number of nodes with degree $k$. Thus, this master equation describes the
introduction of new nodes with degree $m$ and a flux of nodes increasing their 
degrees due to the attachment of new connections. This equation has been originally obtained by Dorogovtsev  and Mendes\cite{Dorogovtsev2000}. The analogous master equations for the number of links $L(k,k',t)$ between nodes of degrees $k$ and $k'$ and number of triangles $\Delta(k,k',k'',t)$ formed by nodes of degrees $k$, $k'$, and $k''$ are
\begin{widetext}
\begin{eqnarray}
\nonumber L(k,k',t+1)=&L(k,k',t)+\delta_{k,m}\frac{m(k'-1)}{2(2mt+N_0k_0)}N(k'-1,t)+\delta_{k',m}\frac{m(k-1)}{2(2mt+N_0k_0)}N(k-1,t)\\
&+\frac{m(k-1)}{(2mt + N_0k_0)}L(k-1,k',t)+\frac{m(k'-1)}{(2mt + N_0k_0)}L(k,k'-1,t)-\frac{m(k+k')}{(2mt + N_0k_0)}L(k,k',t) \; 
\label{master_L}
\end{eqnarray}
and
\begin{eqnarray}
\nonumber
&\Delta(k,k',k'',t+1)=\Delta(k,k',k'',t)+\delta_{k',m}\frac{m^2(m-1)^2(k-1)(k''-1)}{6(2mt+N_0k_0)^2}L(k-1,k''-1,t)
\\
 & +\delta_{k,m}\frac{m^2(m-1)^2(k'-1)(k''-1)}{6(2mt+N_0k_0)^2}L(k'-1,k''-1,t)
+\delta_{k'',m}\frac{m^2(m-1)^2(k-1)(k'-1)}{6(2mt+N_0k_0)^2}L(k-1,k'-1,t)\\
\nonumber & +\frac{m(k-1)}{(2mt +
N_0k_0)}\Delta(k-1,k',k'',t)+\frac{m(k'-1)}{(2mt +
N_0k_0)}\Delta(k,k'-1,k'',t)
+\frac{m(k''-1)}{(2mt + N_0k_0)}\Delta(k,k',k''-1,t)-\frac{m(k+k'+k'')}{(2mt + N_0k_0)}\Delta(k,k',k'',t)  \; .
\label{master_D}
\end{eqnarray}
\end{widetext}

Again, the first terms in these equations, containing the $\delta$-distributions, account for the creation of new links or triangles while the other terms are the evolution of links and triangles due to the change in degree of one participant node.  

For large networks continuous  time 
 may be assumed and Eq. (\ref{master_N}) is
written as a differential equation:
\begin{eqnarray}
\nonumber \frac{\partial N(k,t)}{\partial t} &=& \delta_{k,m} +
\frac{m (k-1)}{(2mt + N_0k_0)}N(k-1,t)\\
&&- \frac{m k}{(2mt + N_0k_0)}N(k,t) \;,
\label{diff_N} 
\end{eqnarray}
representing a set of coupled differential equations, coupling 
each degree $k$ to its predecessor $k-1$. To solve these equations we begin with $k=m$. By our construction,
$N(m-1,t)$ is always zero, and thus the first differential equation  is decoupled:
\begin{equation}
\frac{\partial N(m,t)}{\partial t} = \delta_{k,m} - \frac{m^2}{(2mt +
N_0k_0)}N(m,t) \;.
\end{equation}
This non-homogeneous differential equation, with initial condition
$N(m,0)=0$, can be solved using Green's functions, yielding
\begin{equation}
N(m,t)=\frac{N_0k_0 +2mt}{m(m+2)} -
\frac{(N_0k_0)^{1+m/2}}{m(m+2)(N_0k_0 +2mt)^{m/2}} \; .
\end{equation}
In the limit $t \gg N_0k_0$ the transient term tends to zero and can be neglected 
resulting 
\begin{equation}
N(m,t)=\frac{N_0k_0 +2mt}{m(m+2)} \; .
\end{equation}
This expression is now inserted in the differential equation for
$N(m+1,t)$, what  removes the coupling for the second equation,  and again a solution may be obtained. This process may be repeated \emph{ad infinitum}, and the
expression for each degree may be written as
\begin{equation}
N(k,t)= \frac{(2mt+ N_0k_0)(m+1)}{k(k+1)(k+2)} \; .
\label{sol_N}   
\end{equation}

\begin{figure}
\begin{center}
\includegraphics[width=8cm,clip]{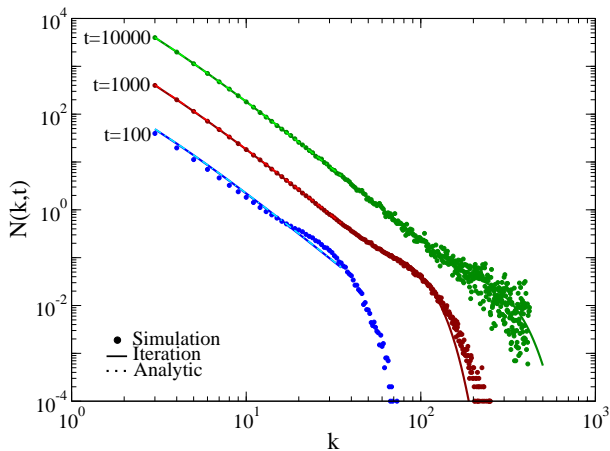} \caption{Number of nodes with degree $k$ for different times. This figure presents a comparison of the analytic solution, the iteration results and simulation data for $N(k,t)$, for different networks sizes, averaged over $10000$ networks. The network size is given by time $t$. Points represent data obtained trough simulation of Barab\'asi-Albert algorithm, the solid lines represent iteration of master equation and dotted lines the analytic solution.}
\label{fig_N}
\end{center}
\end{figure}

\begin{figure}
\begin{center}
\includegraphics[width=8cm,clip]{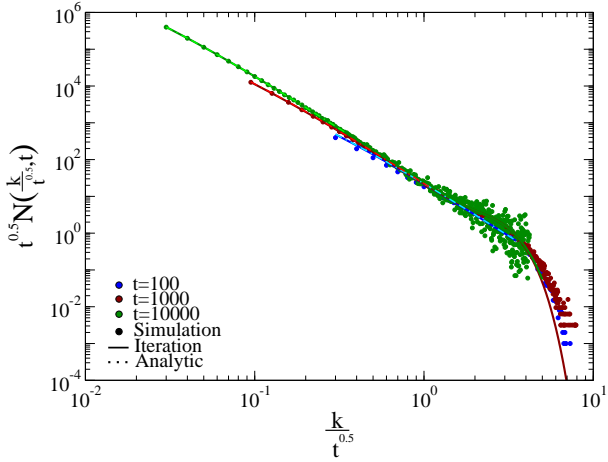}
\caption{Rescaling of $N(k,t)$ by $\sqrt{t}$. This figure presents a rescaling of the number of nodes with degree $k$, comparing the analytic solution, the iteration results and simulation data for $N(k,t)$, for different networks sizes, averaged over $10000$ networks. The network size is given by time $t$. Points represent data obtained trough simulation of Barab\'asi-Albert algorithm, the solid lines represent iteration of master equation and dotted lines the analytic solution. Different times are color coded, blue for $t=100$, red for $t=1000$, and green for $t=10000$.}
\label{fig_N_escala}
\end{center}
\end{figure}

Figure \ref{fig_N} presents  the results from  simulations, iteration of  the master equation, and the analytic solution, Eq. (\ref{sol_N}),  for  the number of nodes with degree $k$ for three different times.  Master equation iteration and the analytic solution fall one above the other, and both agree with simulations for small $k$, deviating only for high values of node degree. This deviation was first described by Dorogovtsev \emph{et. al.} \cite{Dorogovtsev2002} who established $k=\sqrt{t}$ as the upper limit for any analytic description of the model. To verify the limit, Fig.  \ref{fig_N_escala} shows the distribution $N(k,t)$ scaled by $\sqrt{t}$. We can see that the divergence between the analytic solutions and the simulations appears for values $\frac{k}{\sqrt{t}} > 1$. For clarity, the subsequent results will be presented scaled  by $\sqrt{t}$, figures with no scaling can be found in Supplemental Material at [URL].

 Equations (\ref{master_L}) and (\ref{master_D}) can be solved with the same
method applied to Eq.  (\ref{master_N}). Considering continuous time, we  write a set of coupled differential equations associated with  both master equations (\ref{master_L})and ({\ref{master_D}). Solving these equations iteratively, the expression for any set of degrees $k$ and $k'$ are given by (see details in Supplemental Material at [URL]),
\begin{widetext}
\begin{eqnarray}
\nonumber
L(k,k',t)=&\Big[\sum\limits_{i=1}^{k'-m}\binom{k-m+i-1}{i-1}\frac{(m+1)(2mt+N_0k_0)}{2(k'+1-i)(k'+2-i)}\frac{(k-1)!}{(m-1)!}
\frac{(k'-1)!}{(k'-i)!}\frac{(k'+m-i+2)!}{(k+k'+2)!}
+
\\
&\sum\limits_{i=1}^{k-m}\binom{k'-m+i-1}{i-1}\frac{(m+1)(2mt+N_0k_0)}{2(k+1-i)(k+2-i)}\frac{(k'-1)!}{(m-1)!}\frac{(k-1)!}{(k-i)!}\frac{(k+m-i+2)!}{(k+k'+2)!}\Big].
\label{sol_L}
\end{eqnarray}
and
\begin{eqnarray}
\nonumber
\Delta(k,k',k'',t)=&\Big[\sum\limits_{i=0}^{k'-m}\sum\limits_{j=0}^{k''-m}\frac{(k-m+i+j)!}{(k-m-2)!i!j!}\frac{(k-1)!}{(m-1)!}
\frac{1}{6}m(m-1)^2
\frac{(k'-1)!}{(k'-i-2)!}\frac{(k''-1)!}{(k''-j-2)!}\frac{(m+k'-i+k''-j)!}{(k+k'+k'')!}
\frac{L(k'-i,k''-j,t)}{2mt+N_0k_0}
+
\\
\nonumber
&\sum\limits_{i=0}^{k-m}\sum\limits_{j=0}^{k''-m}\frac{(k'-m+i+j)!}{(k'-m-2)!i!j!}\frac{(k'-1)!}{(m-1)!}
\frac{1}{6}m(m-1)^2
\frac{(k-1)!}{(k-i-2)!}\frac{(k''-1)!}{(k''-j-2)!}\frac{(m+k-i+k''-j)!}{(k+k'+k'')!}
\frac{L(k-i,k''-j,t)}{2mt+N_0k_0}
+
\\
&\sum\limits_{i=0}^{k-m}\sum\limits_{j=0}^{k'-m}\frac{(k''-m+i+j)!}{(k''-m-2)!i!j!}\frac{(k''-1)!}{(m-1)!}
\frac{1}{6}m(m-1)^2
\frac{(k-1)!}{(k-i-2)!}\frac{(k'-1)!}{(k'-j-2)!}\frac{(m+k-i+k'-j)!}{(k+k'+k'')!}
\frac{L(k-i,k'-j,t)}{2mt+N_0k_0}\Big].
\label{sol_D}
\end{eqnarray}
\end{widetext}
Observe that in the limit  $t \gg N_0k_0$, Eq. (\ref{sol_L}) is also valid for $k\leq m$ and $k' \leq m$, when the above equation is null. By definition, the
number of links between nodes with degrees $k$ and $k'$ is the same number of
links between nodes with degrees $k'$ and $k$. Therefore $L(k,k',t)$ must be 
symmetrical regarding changes between $k$ and $k'$. This symmetry can be found
in the master equation, Eq. (\ref{master_L}), and its solution Eq. (\ref{sol_L}).
\begin{figure}
\begin{center}
\includegraphics[width=8cm,clip]{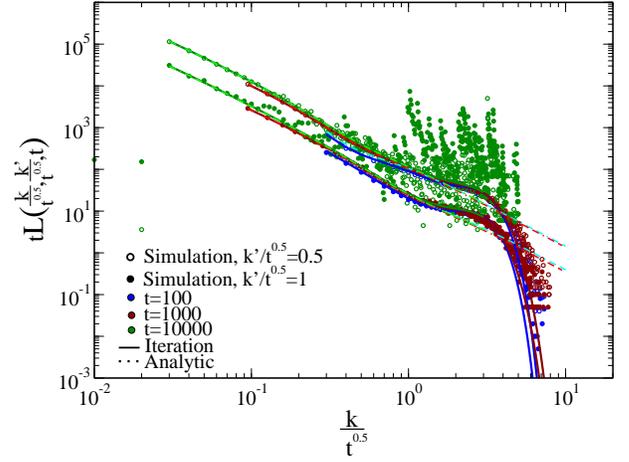}
\caption{Rescaling of number of links between degrees $k$ and $k'$. Bidimensional view of the tridimensional function $L(k,k',t)$ rescaled by $\sqrt{t}$ selecting two values of $k'/\sqrt{t}$, distinguished by solid and empty symbols, and three values of $t$, distinguished by different colors. Simulation data is represented by points, numerical iteration by continuous lines, and analytic solution by dotted lines. Blue symbols and lines represent $t=100$, red symbol and lines represent $t=1000$, and green symbols and lines represent $t=10000$} \label{fig_L}
\end{center}
\end{figure}
Figure \ref{fig_L} presents $tL(k/\sqrt{t},k'/\sqrt{t},t)$ for  $k'=0.5\sqrt{t}$ and $k'=\sqrt{t}$. It is clear that the analytic solution  fits perfectly both simulation and
iteration up to $k \sim \sqrt{t}$. Above this limit the simulation and iteration solutions start deviating from the analytic solutions. 

 Equation (\ref{sol_D}) is  symmetrical relative to 
any change between $k$, $k'$ and $k''$, as expected, given the symmetry of the 
model. Note that this solution is also time independent, since there $L(k,k',t)$ is divided by its time dependence. This occurs because when solving the differential equations  we neglect the transient, and the remaining expression results constant in time. 
This result seems to be counter-intuitive since the model applies for  growing networks. However the dynamics of triangle formation depends also on  link 
distribution, hence being more complex. Therefore the behavior of $\Delta(k,k',k'',t)$ may be elusive. To 
verify Eq. (\ref{sol_D})  we introduce  the one dimensional function $ \rho(r)$  as the number of triangles between nodes with degree $k$, $k'$ and $k''$ with $r=\sqrt{k^2+k'^2+k''^2}$, that is,
\begin{equation}
\rho (r)=\sum\limits_{k}\sum\limits_{k'}\sum\limits_{k''} 
\delta_{r,\sqrt{k^2+k'^2+k''^2}}\Delta(k,k',k'',t) \; .
\end{equation}

\begin{figure}
\begin{center}
\includegraphics[width=8cm,clip]{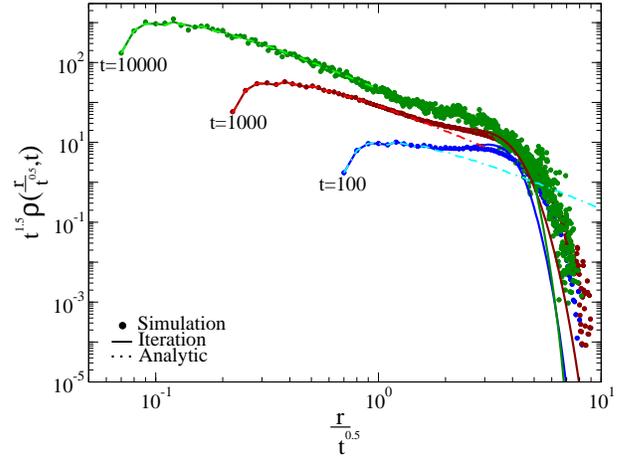} 
\caption{Function $\rho(r)$ rescaled by $\sqrt{t}$. This figure presents three different network sizes and compares simulation, represented by points,  iteration, by solid lines,  and analytic solution, by dotted lines. To compare with different network sizes, the solution was rescaled with the corresponding times.}
\label{fig_D}
\end{center}
\end{figure}
Figure \ref{fig_D} shows $\rho (r)$ scaled by $\sqrt{t}$ . As expected, the 
simulations do not present the same results for different times. However, it is 
possible to note that the curves rapidly converge to the analytic solution for small values 
of $r$. This suggests that the growth dynamic acts on triangles with higher $r$. A 
possible explanation for this behavior lies precisely in the preferential 
attachment mechanism. Since a new node tends to connect to nodes with higher 
degrees, a triangle will be created when two high degree nodes are 
connected to each other. Also, our analytic solution agrees with this 
assumption. Since Eq. (\ref{sol_D}) represents the stationary state, and 
it agrees with the simulation for values of $r$ up to $\sqrt{t}$, it also 
indicates that the number of triangles tends rapidly to a stationary state and any 
difference between Eq. (\ref{sol_D}) and the simulation is given by the 
transient effect of the growth dynamics acting in the most connected nodes.

 In summary, using this method  for solving the differential equations
associated with the discrete time master equations we  obtain analytic
%
solutions for the distributions  of  number of nodes, links and triangles of the
Barab\'asi-Albert model. These results have been verified by numerical iteration of the master equations and simulations, and proved to be valid for small networks,
far from the limit of infinite time. These solutions  describe 
accurately many real networks, failing only for $k > \sqrt{t}$, as predicted by Dorogovtsev \emph{et al.}  In particular, the number of triangles offers a more detailed vision of the neighborhood than average clustering coefficient. Finally, the method employed is general and may 
be applied to different growing network models.

\section*{Acknowledgements}

We acknowledge support from the Centro de F\'\i sica Computacional, Universidade Federal do Rio Grande do Sul. This work has been partially supported by Brazilian agencies FAPERGS, CAPES, and CNPq. It is part of the project PRONEX-FAPERGS 10/0008-0. 


\begin{thebibliography}{10}

\bibitem{Barabasi1999a}
Barab\'{a}si, A. and Albert, R. Emergence of scaling in random networks. \emph{Science} {\bf 286,} 509 (1999).

\bibitem{Barabasi1999b}
Barab\'{a}si, A., Albert, R. and Jeong, H., Mean-field theory for scale-free random networks. \emph{Physica A} {\bf 272,} 173 (1999).

\bibitem{Albert2002}
Albert, R. and Barab\'{a}si, A. Statistical mechanics of complex networks. 
\emph{Rev. mod. phys.} {\bf 74,} 47 (2002).

\bibitem{Barabasi2004}
Barab\'{a}si, A. and Oltvai, Z. Network biology: understanding the cell's functional organization.
\emph{Nat. Rev. Gen.} {\bf 5,} 101 (2004).

\bibitem{Przulj2004}
Przulj, N., Corneil, D., and Jurisica, I. Modeling interactome: scale-free or geometric?
\emph{Bioinformatics} {\bf 20,} 3508 (2004).

\bibitem{Khanin2006}
Khanin, R. and Wit, E., How Scale-Free Are Biological Networks.
\emph{J. Comp. Bio.} {\bf 13,} 810 (2006).

\bibitem{FoxKeller2005}
Keller, E. Revisiting “scale-free” networks. 
\emph{BioEssays} {\bf 27,} 1060 (2005).

\bibitem{Watts1998}
Watte, D. and Strogatz, S. Collective dynamics of 'small-world' 
networks. \emph{Nature} {\bf 393,} 440 (1998).

\bibitem{Ash2007}
Ash, J. and Newth, D. Optimizing complex networks for resilience against cascading failure. \emph{Physica A} {\bf 380,} 673 (2007). 

\bibitem{Eguiluz2002}
Egu\'{\i}luz, V. and Klemm, K. Epidemic Threshold in Structured Scale-Free Networks.
\emph{Phys. Rev. Lett.} {\bf 89,} 108701 (2002).

\bibitem{Newman2003}
Newman, M. The Structure and Function of Complex Networks.
\emph{SIAM Rev.} {\bf 45,} 167 (2003).

\bibitem{Dorogovtsev2000}
Dorogovtsev, S., Mendes, J., and Samukhin, A. Structure of growing networks with preferential linking.
\emph{Phys. Rev. Lett.} {\bf 85,} 4633 (2000).

\bibitem{Dorogovtsev2002}
Dorogovtsev, S. and Mendes, J. Evolution of networks.
\emph{Adv. Phys.} {\bf 51,} 1079 (2002).

\bibitem{Krapivsky2000}
Krapivsky, P., Redner, S. and Leyvraz, F. Connectivity of growing random networks.
\emph{Phys. Rev. Lett.} {\bf 85,} 4629 (2000).

\bibitem{Dunne2002}
Dunne, J., Williams, R. and Martinez, N. Food-web structure and network theory: The role of connectance and size.
\emph{Proc. Natl. Acad. Sci. USA} {\bf 99,} 12917 (2002).

\bibitem{Pastor-Satorras2002}
Pastor-Satorras, R. and  Vespignani, A. Epidemic dynamics in finite size scale-free networks.
\emph{Phys. Rev. E} {\bf 65,} 035108 (2002).

\bibitem{Fraser2005}
Fraser, H. Modularity and evolutionary constraint on proteins.
\emph{Nat. Gen.} {\bf 37,} 351 (2005).

\bibitem{Li2006}
Li, L., Huang, Y., Xia, X., and Sun, Z. Preferential duplication in the sparse part of yeast protein interaction network.
\emph{Mol. Biol. Evol.} {\bf 23,} 2467 (2006).

\bibitem{Ferreira2013}
Ferreira, R. \emph{et al.} Preferential Duplication of Intermodular Hub Genes: An Evolutionary Signature in Eukaryotes Genome Networks.
\emph{PLoS ONE} {\bf 8,} e56579 (2013).

\bibitem{Fotouhi2013}
Fotouhi, B. and Rabbat, M.~G. Network growth with arbitrary initial conditions: Degree dynamics for uniform and preferential attachment.
\emph{Phys. Rev. E} {\bf 88,} 062801 (2013).

\bibitem{Fronczak2003}
Fronczak, A., Fronczak, P. and Hołyst, J. Mean-field theory for clustering coefficients in Barabási-Albert networks.
\emph{Phys. Rev. E} {\bf 68,} 046126 (2003).

\bibitem{Bianconi2003}
Bianconi, G. and Capocci, A. Number of Loops of Size h in Growing Scale-Free Networks.
\emph{Phys. Rev. Lett.} {\bf 90,} 078701 (2003).

\end{thebibliography}
\end{document}